\documentclass[12pt]{article}
\usepackage{setspace}
\usepackage[margin=1in, lines=25]{geometry}
\usepackage{graphicx}
\usepackage{mathrsfs}
\usepackage{natbib}
\usepackage{float}
\usepackage{amsmath, amsthm}
\usepackage{booktabs}
\usepackage{color, caption}
\newcommand{\blue}[1]{{\textcolor{black}{#1}}}

\newtheorem{proposition}{Proposition}
\RequirePackage[colorlinks,citecolor=blue,linkcolor=blue]{hyperref}
\usepackage{bm}
\usepackage{physics}
\newcommand{\pkg}[1]{{\normalfont\fontseries{b}\selectfont #1}}
\renewcommand{\hat}{\widehat}

\renewcommand{\bar}{\overline}
\newcommand{\age}{\textsf{Age}}
\newcommand{\black}{\textsf{Black}}
\newcommand{\married}{\textsf{Married}}

\title{Geographically Weighted Cox Regression for Prostate Cancer Survival Data
in Louisiana}
\author{Yishu Xue\thanks{Department of Statistics, University of Connecticut,
Storrs, CT 06268, USA} ~~ Elizabeth D. Schifano\footnotemark[1] ~~ Guanyu
Hu\footnotemark[1]~\thanks{Correspondence: guanyu.hu@uconn.edu}} \date{\today}
\date{}

\begin{document}
\maketitle

\begin{abstract}
The Cox proportional hazard model is one of the most popular tools in analyzing
time-to-event data in public health studies. When outcomes observed in clinical
data from different regions yield a varying pattern correlated with location, it
is often of great interest to investigate spatially varying effects of
covariates. In this paper, we propose a geographically weighted Cox regression
model for sparse spatial survival data. In addition, a stochastic neighborhood
weighting scheme is introduced at the county level. Theoretical properties of
the proposed geographically weighted estimators are examined in detail. A model
selection scheme based on the Takeuchi's model robust information criteria (TIC)
is discussed. Extensive simulation studies are carried out to examine the
empirical performance of the proposed methods. We further apply the proposed
methodology to analyze real data on prostate cancer from the Surveillance,
Epidemiology, and End Results cancer registry for the state of Louisiana.
\vspace{1ex}

\noindent
{\it Keywords:} Cox Model; Graph Distance; Sparse Spatial Survival Data;
Stochastic Neighborhood Weighting
\end{abstract}

\section{Introduction}\label{sec:intro}

In public health and epidemiology studies, clinical data is often collected at
small geographical levels such as towns and counties, and aggregated on a larger
level such as states. Analysis of such datasets, while providing information on
an overall level about covariate effects, assumes that covariates affect the
outcome equally across different locations, regardless of the variation in local
environment and local treatment. Ignoring these effects makes the estimation on
the aggregated data sub-optimal when we are more interested in precisely
modeling the covariate effects on a finer level. Allowing for spatially related
covariate effects will lead to a more flexible analysis, and a clearer picture
of the relationship between the response variable and the covariates. Several
methods have been proposed for analyzing geographical patterns of survival data.
One popular method is to treat the spatial variation as random effects in the
model. For example, \cite{banerjee2003frailty} considered random effects
corresponding to clusters that are spatially arranged in a frailty model.
\cite{banerjee2005semiparametric} developed a Bayesian hierarchical model that
captures spatial heterogeneity in the framework of proportional odds.
\cite{zhang2011bayesian} added a random effect to the Bayesian accelerated
failure time model with a conditional autoregressive prior to analyze prostate
cancer data from Louisiana collected between 2000 and 2004, while
\cite{li2015spatial} developed a Bayesian semiparametric approach to the
extended hazard model to consider spatial effects on prostate cancer survival.
The aforementioned works only considered spatial random effects on Bayesian
survival models. Another popular approach involves allowing the coefficients of
models to be spatially varying, and using a certain weighting scheme in
estimating the coefficients for different locations.
\cite{gelfand2003spatial}
considered spatially varying coefficients in the context of Gaussian responses,
and proposed a Gaussian process model to estimate the coefficients.
\cite{brunsdon1998geographically} proposed an alternative approach where a
weighting function, based on a certain measure of distance, is imposed on the
observations, and the parameter estimates can be obtained from a weighted least
squares estimation. \cite{nakaya2005geographically} extended the idea to Poisson
regression, and the parameter estimates are obtained by maximizing a weighted
likelihood function.


For survival data with time-to-event structure, \cite{hu2018modified} studied a
spatially varying Nelson--Aalen estimator and Kaplan--Meier estimator with a
geographically weighted estimating approach. \cite{hu2017spatial} studied an
accelerated failure time regression model with spatially varying coefficients in
a Bayesian context. For the Cox proportional hazards
model~\citep{cox1972,cox1975partial}, \cite{fan2006local} studied time varying
coefficients using a local partial-likelihood estimator. Geographically weighted
survival models, however, are not fully studied in spatial survival analysis,
and there is no existing literature studying the Cox model with spatially
varying coefficients for geographically distributed survival data.

For geographically sparse data, simple data stratification by location (e.g., by
county) is not feasible as the sample sizes may be too small to fit Cox models
for some locations. In order to address this issue, we develop a geographically
weighted Cox regression model following the first law of
geography~\citep{tobler1970computer}. Similar to the idea from
\cite{brunsdon1996geographically,brunsdon1998geographically}, we estimate the
regression
coefficients
on
each individual location (i.e., county) by maximizing the local
partial-likelihood with subjects weighted according to their distance from this
location. Greater weights are assigned to the nearby subjects, since
intuitively, nearby regions will share similar environmental and social factors,
which will have similar effects on the observations.

\blue{As with most geographically weighted methods, as recently reviewed in
\cite{murakami2019importance}, the choice of weighting function, its associated
bandwidth, and distance metric are all important. 
In the geographically weighted regression (GWR) context, 
\cite{brunsdon1996geographically} discussed different weighting functions,
and the bandwidth is selected by minimizing the cross-validated out-of-sample
sum of squared errors on a grid of candidate values. 
The choice of distance metric in GWR has been discussed by
\cite{lu2019response,oshan2019comment}.
\cite{white2009stochastic}
used a Stochastic Neighborhood Autoregressive (SNCAR) model where observations
that are beyond a certain threshold of distance are weighted downward. The
choice
of such threshold can be subtle, and can impact the 
final model estimation results. To overcome such
complication, we consider the usage of graph distance
\citep{muller1987algorithm,bhattacharyya2014community} which yields a natural
choice of threshold, provides robust estimation, and can be easily
implemented. 
While cross-validation remains a popular approach for choosing
bandwidths in the linear regression framework, prediction-based selection
methods
are not suitable in this scope as there is no response for the hazard, and only
the
parametric component of the hazard function is estimated by Cox regression using
the partial likelihood approach.
 Therefore, a
likelihood-based bandwidth selection approach is proposed. To account for the
bias-variance tradeoff in addition to maximizing the partial likelihood, a
modified Takeuchi information criterion \citep[TIC;][]{Takeuchi:1976fb} is used
in favor of the Akaike information criterion
\cite[AIC;][]{akaike1973information}. }


The rest of this paper is organized as follows. In Section~\ref{sec:motiv}, data
from the Surveillance, Epidemiology, and End Results (SEER) cancer registry is
introduced as a motivating example. In Section~\ref{sec:gwCox}, we propose the
geographically weighted Cox model, and modify the stochastic neighborhood
weighting function of \cite{white2009stochastic} for areal based data using the
graph distance. \blue{Model selection based on TIC, as well as the theoretical
properties of the estimators, are discussed in the same section.} Simulation
studies to illustrate the performance of our estimators
are presented in Section~\ref{sec:sim}. In Section~\ref{sec:realdata}, we apply
our methods to survival analysis of patients diagnosed with prostate cancer from
the SEER cancer registry for the state of Louisiana. We conclude the paper with
a brief discussion in Section~\ref{sec:disc}.

\section{Motivating Example}\label{sec:motiv}

The SEER Program provides information on cancer statistics in an effort to
reduce the cancer burden among the U.S. population. We consider the prostate
cancer data from July to December 2005 diagnoses for Louisiana from their
November 2014 submission~\citep{hu2018modified}.

Due to Hurricane Katrina, the number of observations diagnosed between the
second half of 2005 is noticeably smaller than for other years (1403 vs 1787 for
July to December for years 2000-2012 except for 2005). The data being spatially
sparse, these diagnoses are not analyzed in most SEER reportings. Except for the
relatively small sample size, censoring rates and other descriptive measures of
covariates presented in Table~\ref{tab:demo} are similar to the 2000
to 2004 dataset discussed in~\cite{zhang2011bayesian}.

\begin{figure}[tbp]
\centering
\includegraphics[width = \linewidth]{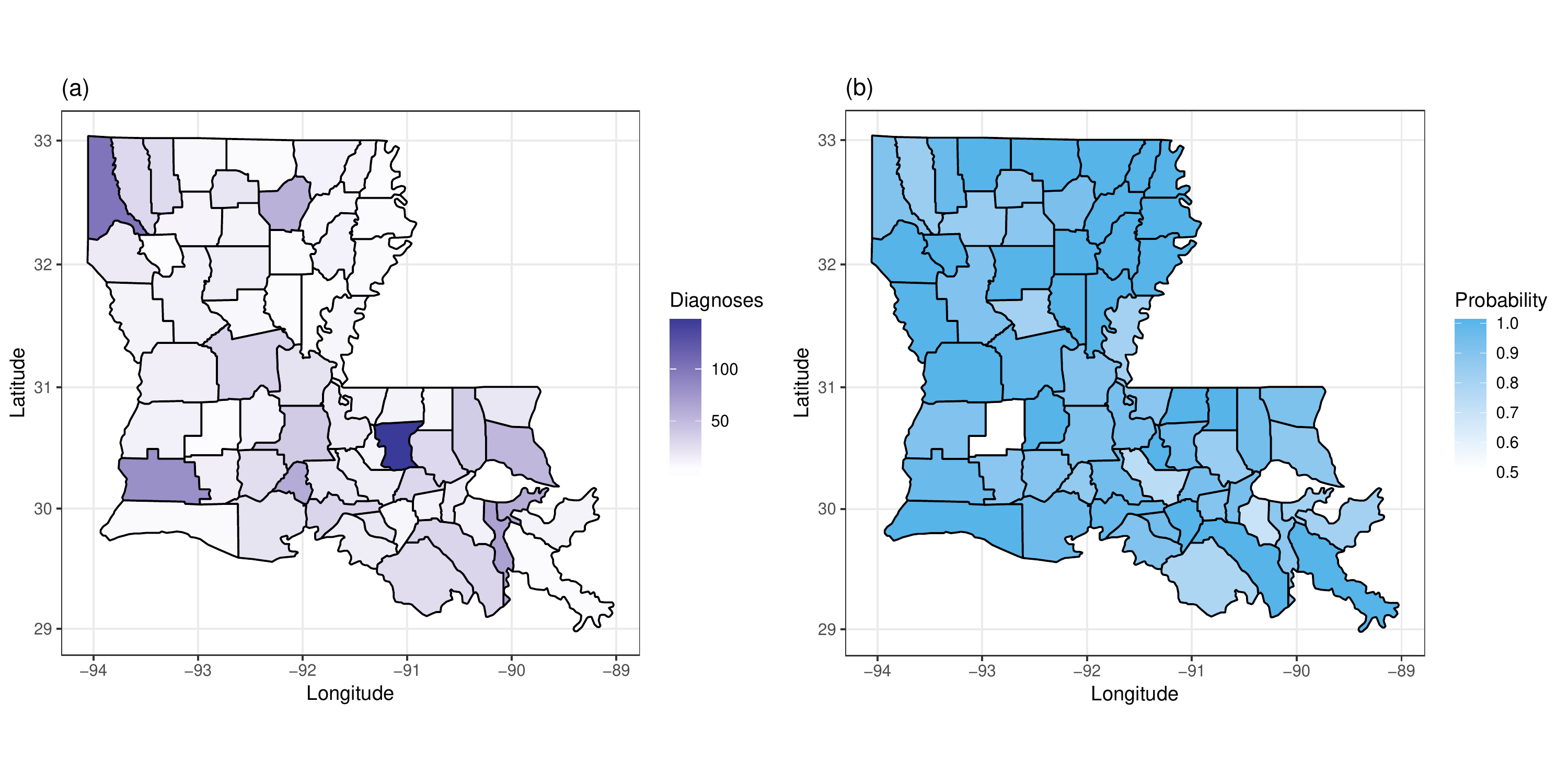}
\caption{(a) Number of diagnoses in counties of Louisiana; (b) Kaplan--Meier
estimate of survival probability at 50 months after
diagnosis.}\label{fig:popplot}
\end{figure}

Together with the covariates, survival times, final statuses, and county
locations of these observations are also reported. Only events due to prostate
cancer are considered. Figure~\ref{fig:popplot} presents the number
of diagnoses, and the Kaplan--Meier estimate of survival probability at 50
months after diagnosis in the 64 counties of Louisiana. The distributions of the
covariates from different counties are very similar. The Kaplan--Meier estimate
of survival probability, however, shows a spatially varying pattern across
counties, which is similar to observations made in \cite{hu2018modified}. In
order to capture the spatial heterogeneity of the hazard rate, we consider the
Cox model with spatially varying coefficients. 

\begin{table}[tbp]
\centering
\caption{Demographic characteristics for Louisiana data. For continuous
variables, the mean and standard deviation (SD) are reported. For binary
variables, the frequency and percentage of each class are reported.}
\begin{tabular}{lc}
\toprule
 & Mean (SD)/ Frequency (Percentage) \\ \midrule Age & 66.21 (11.25) \\
Survival Time & 72.36 (25.54) \\
~~Event  & 32.41 (24.17) \\
~~Censor & 76.01 (22.35) \\
Marital Status \\
~~Currently Married & 857 (67.11\%) \\
~~Other & 420 (32.89\%) \\
Race \\
~~Black & 379 (29.68\%) \\
~~Other & 898 (70.32\%)  \\
Cause-specific Death Indicator \\
~~Event & 107  (8.38\%)\\
~~Censor & 1170 (91.62\%) \\ \bottomrule
\end{tabular}
\label{tab:demo}
\end{table}

\section{Methodology}\label{sec:gwCox}

\subsection{Geographically Weighted Cox Model}

We first consider the case where the precise location of each observation is
available, and is represented in (latitude, longitude) pairs. Let
$(T_i,\delta_{i},Z_i, s_i)$, $i=1,...,n$, denote an independent sample of
right-censored survival data from different sites, where $T_i$ is a
right-censored event time and $\delta_i = I(T_i^* \leq  C_i)$ with $T_i^*$ being
the true survival time and $C_i$ being the censoring time, $Z_i \in
\mathcal{R}^p$ is the corresponding vector of covariates, and $s_i \in
\mathcal{R}^2$ is the corresponding location, i.e., $s_i =
(\text{latitude}_i,\text{longitude}_i)$. In the proportional hazards model,
estimation of $\beta$ is achieved via maximization of the partial likelihood
\citep[using notation of][]{fleming1991}:
\begin{equation}\label{eq:plklhd}
\mathrm{PL}(\beta) = \prod_{i=1}^n \prod_{t\geq 0} \left[
\frac{Y_i(t)\exp\left(Z_i^\top\beta\right)}
{\sum_{j} Y_j(t)\exp\left(Z_j^\top\beta\right)}\right]^
{\dd N_i(t)},
\end{equation}
where $Y_i(t) = I(T_i \geq t)$, $N_i(t)\in\{0,1\}$ is the count of events for
subject $i$ at time $t$, and $\dd N_i(t) = I(T_i\in [t, t + \Delta), \, \delta_i
= 1)$, where $\Delta$ is chosen to be very small such that $\sum_{i=1}^n \dd
N_i(t) \leq 1$ for any $t$.

This estimation scheme essentially assigns equal weights to all observations. As
mentioned in the introduction, estimating regression coefficients for each
county in the Cox model by using only observations within that county is often
not feasible, as in many cases (e.g., for relatively rare diseases) there are
not enough observations from one particular county to fit the model. We thus
propose a geographically weighted Cox model, inspired from the work of
\cite{brunsdon1998geographically}, using a modified local partial likelihood at
a general location $s$ is given by
\begin{equation}\label{eq:plklhd_local}
\mathrm{PL}(\beta(s)) = \prod_{i=1}^n \prod_{t\geq 0} \left[
\frac{w_i(s)Y_i(t)\exp\left\{Z_i^\top\beta(s)\right\}}
{\sum_{j} w_j(s)Y_j(t)\exp\left\{Z_j^\top\beta(s)\right\}}\right]^
{\dd N_i(t)},
\end{equation}
where $w_i(s)$ is the geographical weight calculated using the distance between
$s$ and $s_i$. The choice of weight and distance measure will be discussed in
Section~\ref{ssec:weighting} below. Setting the first order derivatives of $\log
\mathrm{PL}(\beta(s))$,
\begin{equation}\label{eq:plklhd_first}
\begin{split}
\frac{\partial \log
\mathrm{PL} (\beta(s))}{\partial \beta(s)}
= \sum_{i=1}^n \delta_i\left(w_i(s)
Z_{i}-\frac{\sum_{j \in
R(T_i)}w_j(s)Z_{j}\exp\left\{Z_{j}^\top \beta(s)\right\}}{\sum_{j
\in R(T_i)}w_j(s)\exp \left\{Z_{j}^\top \beta(s)\right\}}\right),
\end{split}
\end{equation}
to zero yields estimates of $\beta(s)$ using the Newton-Raphson technique, where
$ R(T_i)$ is the set of observations with $Y_j(T_i)=1$. We can also obtain the
observed Fisher information of~\eqref{eq:plklhd_local}:
\begin{equation}\label{eq:weightedinfo}
\begin{split}
    \mathcal{I}(\beta(s)) = \sum_{i=1}^n\left[ \frac{\sum_{j\in
\mathcal{R}(T_i)}w_j(s)^2[\exp\{Z_{j}^\top\beta(s)\}]^2Z_{j}Z_{j}^\top}{\left
[\sum_{j\in\mathcal{R}(T_i)} w_j(s)\exp\{Z_{j}^\top\beta(s)\}\right]^2} -
\frac{\sum_{j\in\mathcal{R}(T_i)}w_j(s)\exp\{Z_{j}^\top\beta(s)\}Z_{j}
Z_{j}^\top}{\sum_{j\in\mathcal{R}(T_i)} w_j(s)\exp\{Z_{j}^\top\beta(s)\}}
\right].
\end{split}
\end{equation}

\subsection{Stochastic Neighborhood Weighting Function}\label{ssec:weighting}

We now review some traditional weighting schemes for geographically weighted
regression.  Suppose again, for now, that the precise (latitude, longitude)
location of each observation is available. As in \cite{hu2017spatial} and
\cite{hu2018modified}, a natural way to account for the locality is setting the
weights to zero if the location of the observation exceeds some threshold $d$
from the location whose vector of coefficients we want to estimate. This induces
the weighting scheme:
\begin{equation}\label{eq:disweight}
    w_i(s)= \begin{cases}
        1 & d_i(s) < d \\
        0 & \text{otherwise}
    \end{cases},
\end{equation}
where $d_i(s)$ is a certain measure of distance between locations $s_i$ and $s$.
These weights are the simplest to calculate, but are discontinuous as a function
of the distance between the two locations. Alternatively, we may use the
exponential function or  Gaussian function to compute continuous weights:
\begin{equation*}
  \begin{array}{ll}
w_i(s)=\exp(-d_i(s)/h) & \text{(exponential)}\\
w_i(s)=\exp(-(d_i(s)/h)^2) & \text{(Gaussian)}
  \end{array}
\end{equation*}
where $h>0$ is a bandwidth parameter chosen by the user. Both weighting
functions are decreasing with respect to the distance between two locations.
\blue{The bi-square kernel, which takes the form
\begin{equation*}
    w_i(s) = \begin{cases}
        1 - (d_i(s) / d) ^2 & |d_i(s)| < d \\
        0 & \mbox{otherwise}
    \end{cases},
\end{equation*}
again with $d$ being some threshold, has also been used in many works,
including \cite{brunsdon1996geographically} and \cite{oshan2019mgwr}.}

The weighting functions mentioned above are mainly appropriate for
point-reference data where locations vary continuously over a spatial domain.
The data considered in our study is areal data, where the spatial domain is a
fixed subset (of regular or irregular shape), but now partitioned into a finite
number of areal units (e.g., counties) with well-defined boundaries. One natural
way to proceed is to locate all observations for a county to its ``center'', for
example, its centroid, and calculate a certain measure of distance between the
county centroids, e.g., the great circle distance \citep{Rpkg:geosphere},
\blue{or
the Euclidean distances calculated based on the projected coordinates of
county centroids \citep{oshan2019mgwr}.}
Then
based on such distance matrices, \cite{white2009stochastic} proposed the
SNCAR model, which is an extension of the ordinary Conditional Autoregressive
\citep[CAR;][]{banerjee2014hierarchical}
model. Unlike the general adjacency matrix, whose diagonal elements are all 0
and off diagonal element $a_{ij}=1$ if areas $A_i$ and $A_j$ share a common
boundary, the SNCAR model allows the off-diagonal elements to depend on unknown
parameters, i.e., 
\begin{equation}
    a_{ij}=\begin{cases}
    1& \text{if } 0<d_{ij}\leq d_l\\
    c(d_{ij},h)& \text{if } d_l<d_{ij}
    \end{cases},
    \label{eq:stochastic_neighbor}
\end{equation}
where $c(d_{ij},h)$ is some function such that $c(d_{ij},h)<1$ (e.g.,
$c(d_{ij},h)=\exp(-d_{ij}/h)$ ), $d_l$ is an unknown \blue{threshhold} for
adjacency
to be estimated, and $d_{ij}$ is a certain measure of distance between $A_i$ and
$A_j$. Such adjacency matrices can subsequently be used to assign weights to
observations in different counties when we fit regression models for each
particular county. It can be seen that the choice of distance function, as well
as the distance measure, are both critical in deciding the weights. Deciding on
a threshold $d_l$ \blue{in \eqref{eq:stochastic_neighbor}} for the great circle 
distance \blue{or Euclidean distance,} for example, involves
determining how close is ``close enough" to be considered the same, which is
usually a subjective matter. In practice, it is often done with cross-validation
such as in \cite{brunsdon1998geographically}, which is highly data dependent. A
more robust and natural distance function, as well as an associated rule, is
desired. Therefore, as a solution to this problem, we propose the usage of the
graph distance \citep{bhattacharyya2014community} in formulating adjacency
matrices.

\begin{figure}[tbp]
\centering
\includegraphics[width = \textwidth]{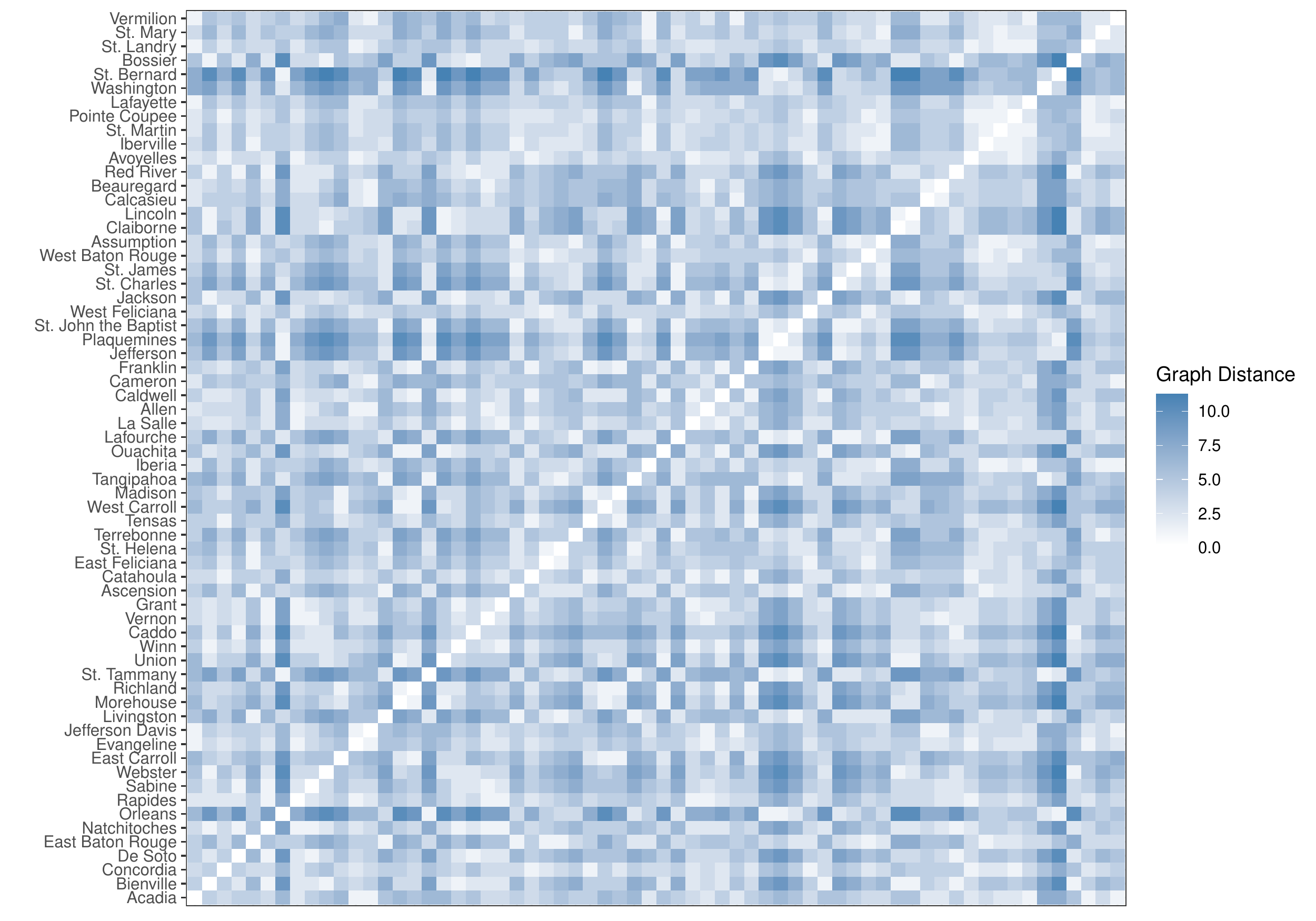}
\caption{Visualization of graph distances for Louisiana counties. Darker colors
indicate greater graph distances. The maximum graph distance is 11.}
\label{fig:distmat}
\end{figure}

Following \cite{muller1987algorithm} and \cite{bhattacharyya2014community}, we
denote a graph as $G$, with set of vertices $V(G)=\{v_1,\ldots, v_n\}$, and set
of edges $E(G) = \{e_1,\ldots, e_m\}$. The graph distance between two vertices
$v_i$ and $v_j$ is defined as follows:
\begin{equation}
d_{v_i v_j}=\begin{cases}
|V(e)| &\text{if } e \text{ is the shortest path connecting } v_i
\text{ and } v_j \\
\infty & v_i \text{ and }\, v_j \text{ are not connected}
\end{cases},
\label{eq:netdistance}
\end{equation}
where $|V(e)|$ represents the cardinality of edges in $e$. In this way, we can
calculate the graph distance among the counties  in the data set. In other
words, we treat the spatial structure of Louisiana as a graph, and each county
is represented as one vertex of this graph. The distance matrix of the 64
Louisiana counties in our study is provided in
Figure~\ref{fig:distmat} for illustration. Plugging in the graph
distances into \eqref{eq:stochastic_neighbor} yields a weighting function to
assign a weight to observations in county $i$ when we fit a regression model for
county $s$:
\begin{equation}
w_i(s)=\begin{cases} 1& \text{ if } d_{v_i v_s} \leq 1\\
    \exp(-d_{v_i v_s} / h) & \text{ if } 1 < d_{v_i v_s}
\end{cases},
\label{eq:stochastic_neighbor_weights}
\end{equation}
where $d_{v_iv_s}$ is the graph distance between counties $i$ and $s$. This
means that all observations in county $i$ will get the same weight value. The
spatial weighting function in \eqref{eq:stochastic_neighbor_weights} will use
all the information from adjacent counties, and the weight is a decreasing
function of the between-county distance.

\subsection{Model Assessment Criterion}

The bandwidth $h$ in \eqref{eq:stochastic_neighbor_weights} should not be chosen
arbitrarily. For Cox regression on a dataset where covariates are fixed but
their effects are time-varying, \cite{verweij1995time} suggested using partial
likelihood based \blue{AIC for model selection. Here, as we are only concerned
with
bandwidth selection, the component of AIC accounting for model complexity
remains the same across all models being compared, so that the
log-partial likelihood would be the sole determinant of AIC values. The smallest
bandwidth would always be preferred, as the model is driven to fit to local data
as closely as possible. The
bias-variance tradeoff in introducing bias, while at the same time lowering
volatility of per-county parameter estimates, is not taken into account.
Therefore, we consider using the TIC.} Using the generalized dimension of a
model
instead of simply the dimension of parameters, the TIC for a general model $M_l$
is defined as:
\begin{equation}
\text{TIC}(M_l)=\max_{\theta_l}\left[ -2\log (\mathcal{L}(M_l,\theta_l))-
2\text{Tr}(\mathcal{I}^{-1}(\theta_\ell)K(\theta_\ell))\right],
\end{equation}
where $\mathcal{L}$ denotes the model likelihood, $\theta_l$ is the parameter
in model $M_l$, $\mathcal{I}(\theta_\ell)$ is the information matrix for
$\theta_\ell$, and $K(\theta_\ell) = U(\theta_\ell)U^\top(\theta_\ell)$
with
$U(\theta_\ell)$ being the score vector at~$\theta_\ell$.

We consider the partial likelihood based TIC for the geographically weighted Cox
model. For observed data with $J$ unique locations $s^*_1,\cdots,s^*_J$ and
estimated regression coefficients
$\hat{\beta}(s^*_1),\cdots,\hat{\beta}(s^*_J)$, for a particular bandwidth $h$,
the TIC can be defined as:
\begin{equation}
\begin{split}
\text{TIC}(h)=&-2\sum_{j=1}^{J} \sum_{1\leq i \leq n, s_i=s_j^*}
\delta_i \left\{Z_i^\top\hat{\beta}(s_j^*)-\log\left[\sum_{k \in
\mathcal{R}(T_i)}\exp\left(Z_k^\top\hat{\beta}(s_j^*)\right
)\right]\right\} \\
&+2\sum_{j=1}^J \text{Tr}\left( \mathcal{I}^{-1}(\hat{\beta}(s_j^*))
K_j(\hat{\beta}(s_j^*))\right),
\label{eq:TIC_b}
\end{split}
\end{equation}
where $\mathcal{I}^{-1}(\hat{\beta}(s_j^*))$ is the observed information matrix
in~\eqref{eq:weightedinfo} evaluated at $\hat{\beta}(s_j^*)$, and
$K_j(\hat{\beta}(s_j^*)) = U_j(\hat{\beta}(s_j^*))U_j(\hat{\beta}(s_j^*))^\top$
is the variance matrix for the score vector based on observations at $s_j^*$,
which can be calculated as:
\begin{equation*}\label{eq:unweightscore}
U_j(\hat{\beta}(s_j^*)) = \sum_{i=1}^{n_j} \int_0^\infty [Z_i -
\bar{Z}(\hat{\beta}(s_j^*),t)]
\dd N_i(t),
\end{equation*}
where $n_j$ is the number of observations at $s_j^*$, and
$\bar{Z}(\hat{\beta}(s_j^*),t)$ is the weighted vector of covariates of these
observations still alive at time $t$, with the weights being their risk scores,
$\exp\{Z_i^\top \hat{\beta}(s_j^*)\}$. We can select the bandwidth with the
smallest TIC calculated by \eqref{eq:TIC_b}. In our Louisiana dataset, the
$s_j^*$'s correspond to the centroids of the 64 counties.

\subsection{Asymptotic Results}

The following regularity conditions are required to establish consistency of the
estimator and the asymptotic distribution of the estimator:
\begin{enumerate}
\item [C1] Conditions A1 to A8 in \cite{fan2006local}.
\item [C2] $w_i(s)$, $i=1,\cdots,n$ follow regularity conditions of Theorem 3.2
in \cite{wang2006estimation}.
\end{enumerate}

Condition C1 and C2 will be used to derive the pointwise convergence properties
of $\hat{\beta}(s)$ and its asymptotic normality. Condition C2 will be used to
derive the bias term of $\hat{\beta}(s)$ in Proposition 2 below and has
regularity constraints on the geographical weights.  The proofs are provided in
the Supplemental Material.

\begin{proposition}
Under condition C1, we have:
\begin{equation*}
\hat{\beta}(s)\overset{p}{\longrightarrow} \beta(s),
\end{equation*}
for any $s \in \mathcal{R}^2$.
\label{prop1}
\end{proposition}

\begin{proposition}
Under conditions C1 and C2, we have:
\begin{equation*}
\sqrt{n}(\hat{\beta}(s)-\beta(s)-\xi(s)
)\overset
{\mathcal{L}}{\longrightarrow} N(0,\Sigma(s)),
\end{equation*}
for any $s \in \mathcal{R}^2$, where $\xi(s)$  is bias of $\hat{\beta}(s)$
defined as $b_{nw}$ in Theorem 3.2 in \cite{wang2006estimation}, and $\Sigma(s)$
is the asymptotic variance covariance matrix.
\label{prop2}
\end{proposition}

\section{Simulation Study}\label{sec:sim}

In this section, we use simulated survival datasets that resemble the SEER data
to study the performance of the proposed method when the observations are
generated with and without spatially varying coefficients. All
calculations are performed in \textsf{R}, using the \pkg{survival}
package~\citep{therneau2017}. The code and related documentation
is available at GitHub.
\blue{Additional simulation studies
have been conducted to verify the performance of the proposed methods for
high censoring survival data, and the results are included in the Supplemental
Material}.

\subsection{Simulation Without Spatially Varying Coefficients}

We obtain the geographical information of Louisiana counties from the US Census
Bureau. The same spatial correlation structure of counties from the real data
is used in our simulation. In addition, to compare geographical distance based
weighting schemes, the matrix of great circle distance \citep{Rpkg:geosphere}
for the 64 county centroids is also calculated. As the maximum graph distance
for Louisiana counties is 11, the great circle distances are also normalized to
have a maximum value of 11 to make both distances comparable.

To begin, the number of observations for each county is randomly selected from
30 to 40 to give an expected sample size of 35 per county. The sample size is
selected to be slightly larger than the average per county sample size in the
real data, as we would like to obtain the local parameter estimates as well for
comparison. Three covariates are considered: \age, (centered and scaled),
\black, and \married, where $\age \sim N(0,1)$, $\black \sim
\text{Bernoulli}(0.3)$, and $\married \sim \text{Bernoulli}(0.7)$. Next,
survival times are generated from a Cox model with baseline hazard $\lambda_0(t)
= 0.03$, and vector of coefficients $\beta = (0.7, 0.5, -0.8)^\top$. Censoring
times are generated independently using a mixture distribution
$0.1\text{Uniform}(0, 60) + 0.9 \langle 60 \rangle$, where $0.1 \langle 60
\rangle$ represents a point mass at 60. The average censoring rate is around
40\%. Weights based on the graph distance matrix
(Figure~\ref{fig:distmat}) are calculated using the weighting scheme
in~\eqref{eq:stochastic_neighbor_weights}. For the great circle distance,
\eqref{eq:stochastic_neighbor} is used with $c(d_{ij,j}) = \exp(-d_{ij}/h)$ and
four candidate values for $d_l$: 0.5, 1, 2, and 1.29. The last threshold,
$d_l=1.29$ corresponds to the great circle distance based weight matrix which
has same number of 1 entries as the graph distance based matrix
from~\eqref{eq:stochastic_neighbor_weights}.

With the largest distance being 11, at $h=50$, a county will have a relative
weight of over 0.80 in estimating even the most distant county, and the
performance of such a model is fairly close to a global model where all
observations are used and weighed equally. Therefore, the grid of bandwidths
considered is set to $h\in\{0.5, 1, \ldots, 50\}$. For each county, we fit the
geographically weighted Cox regression using the different aforementioned
weighting schemes, a global (unweighted) Cox regression using all observations,
and a local Cox regression using only observations within the particular county.
The simulation process described above is repeated for $r=1000$ times. The
parameter estimates are evaluated using the following four measurements:
\begin{align*}
    \text{mean absolute bias (MAB)} &=  \frac{1}{64}\sum_{\ell=1}^{64}
\frac{1}{1000}\sum_{r=1}^{1000} \left|\hat{\beta}_{\ell,m,r} -
\beta_{\ell,m}\right|,
\\
\text{mean standard deviation (MSD)} & = \frac{1}{64}\sum_{\ell=1}^{64}
\sqrt{\frac{1}{999}\sum_{r=1}^{1000} \left(\hat{\beta}_{\ell,m,r} -
\bar{\hat{\beta}}_{\ell,m}\right)^2}, \\
\text{mean of mean squared error (MMSE)} & =  \frac{1}{64}\sum_{\ell=1}^{64}
\frac{1}{1000}\sum_{r=1}^{1000} \left(\hat{\beta}_{\ell,m,r} -
\beta_{\ell,m}\right)^2,\\
\text{mean coverage probability (MCP)} &=
\frac{1}{64}\sum_{\ell=1}^{64}\frac{1}{1000}
\sum_{r=1}^{1000}1\left(\left|\hat{\beta}_{\ell,m,r} - \beta_{\ell,m}
\right| \leq 1.96 \text{SE}(\hat{\beta}_{\ell,m,r} )\right),
\end{align*}
where $\hat{\beta}_{\ell,m,r}$ is the estimate for the $m$th coefficient of
county $\ell$ in the $r$th replicate, $\overline{\hat{\beta}}_{\ell,m}$ is the
average of $\hat{\beta}_{\ell,m,r}$ over the 1000 replicates, $\beta_{\ell,m}$
is the true underlying parameter, and $1(\cdot)$ is the indicator function.

\begin{figure}[tbp]
	\centering
	\includegraphics[width = .8\textwidth]{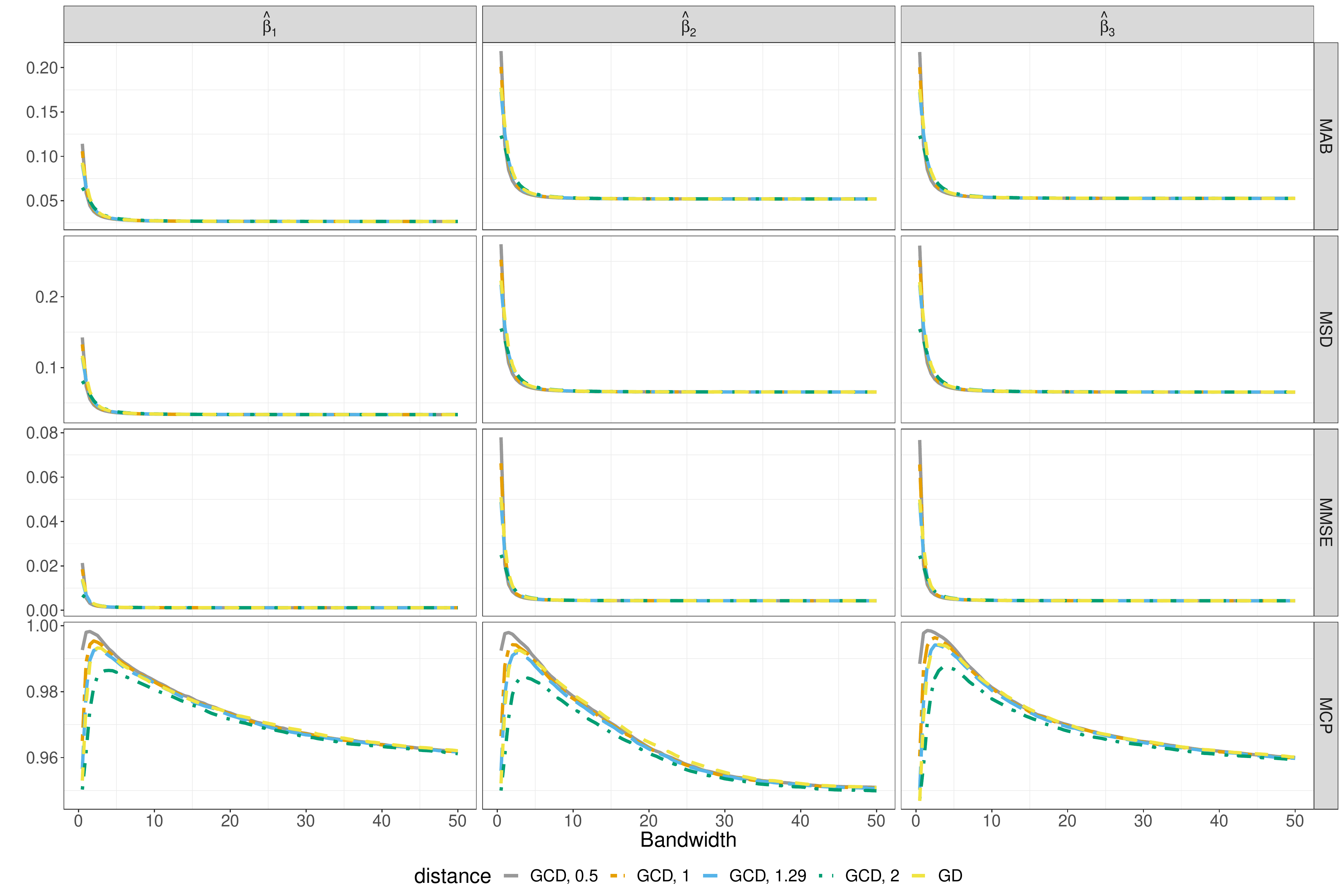}
	\caption{Performance measures for the geographically weighted Cox models
fitted using different distance/threshold combinations when there is no
spatial variation in covariate effects.}
	\label{fig:nullplot}
\end{figure}

The four measurements are plotted against $h$ in Figure~\ref{fig:nullplot}. As
seen from the graph, when there is no spatial variation in covariate effects, at
large bandwidth values, the MAB, MSD and MMSE stabilize at very small values,
while the MCP stays well above 0.95. The most frequently selected bandwidth by
the graph distance based weighted models, i.e., the bandwidth that corresponds
to the smallest TIC in the 1000 replicates, is 50, which is in accordance with
our expectation. With $d_l=0.5$ and $d_l=1$ in the great circle distance based
models, the TIC is dominated by the likelihood component, and tends to select
the smallest bandwidth possible ($h=0.5$), where little neighboring information
is take into account. The MAB, MSD and MMSE, however, are fairly large at this
bandwidth. With $d_l=1.29$ and $d_l=2$, the great circle distance based models
favor the largest bandwidth possible, and their performances are highly similar
to the graph distance based model, as they all approximate a globally unweighted
model.

\begin{table}[tbp]
\centering 
\caption{Performance of local, global, and best selected geographically
weighted Cox regressions when there is no spatial variation in covariate
effects. GD stands for graph distance, and GCD stands for great circle
distance.}
\label{tab:null}
\begin{tabular}{lcccccc}
\toprule
	Model & Parameter & MAB & MSD & MMSE & MCP\\
	\midrule 
	Local & $\beta_1$ &   0.299 & 0.393 & 0.165 & 0.946      \\
	      & $\beta_2$ &     0.586 & 1.163 & 1.409 & 0.944     \\
          & $\beta_3$ &     0.568 & 0.979 & 1.019 & 0.944    \\ [.5ex]
	Global & $\beta_1$ &   0.027 & 0.034 & 0.001 & 0.949     \\
	      & $\beta_2$ &   0.052 & 0.065 & 0.004 & 0.945   \\
          & $\beta_3$ &    0.053 & 0.065 & 0.004 & 0.953   \\ [.5ex]
	GD Weighted, $h=50$ & $\beta_1$ &   0.027 & 0.034 & 0.001 & 0.962    \\
	      & $\beta_2$ &    0.052 & 0.066 & 0.004 & 0.951        \\
          & $\beta_3$ &    0.053 & 0.065 & 0.004 & 0.960      \\ [.5ex]
	GCD Weighted, $d_l=0.5, h=0.5$ & $\beta_1$ &  0.114 & 0.143 & 0.021 & 0.993 \\
	      & $\beta_2$ & 0.219 & 0.274 & 0.078 & 0.992        \\
          & $\beta_3$ & 0.218 & 0.272 & 0.077 & 0.988  \\[.5ex]
	GCD Weighted, $d_l=1, h=0.5$ & $\beta_1$ & 0.106 & 0.133 & 0.019 & 0.965 \\
	      & $\beta_2$ &  0.201 & 0.252 & 0.066 & 0.967    \\
          & $\beta_3$ &  0.200 & 0.251 & 0.066 & 0.960      \\ [.5ex]
	GCD Weighted, $d_l=1.29, h=50$ & $\beta_1$ & 0.027 & 0.034 & 0.001 & 0.962 \\
	      & $\beta_2$ & 0.052 & 0.066 & 0.004 & 0.951     \\
          & $\beta_3$ & 0.053 & 0.065 & 0.004 & 0.960    \\ [.5ex]
	GCD Weighted, $d_l=2, h=50$ & $\beta_1$ &  0.027 & 0.034 & 0.001 & 0.961 \\
	      & $\beta_2$ &  0.052 & 0.066 & 0.004 & 0.950        \\
          & $\beta_3$ &   0.053 & 0.065 & 0.004 & 0.959      \\
\bottomrule
\end{tabular}
\end{table}

\subsection{Simulation with Spatially Varying Coefficients}

To investigate the performance of the proposed methods in the presence of
spatially varying coefficients, the same procedures as previously described
are used to generate the covariates and censoring times. 

\begin{figure}[tbp]
	\centering
	\includegraphics[width = .8\textwidth]{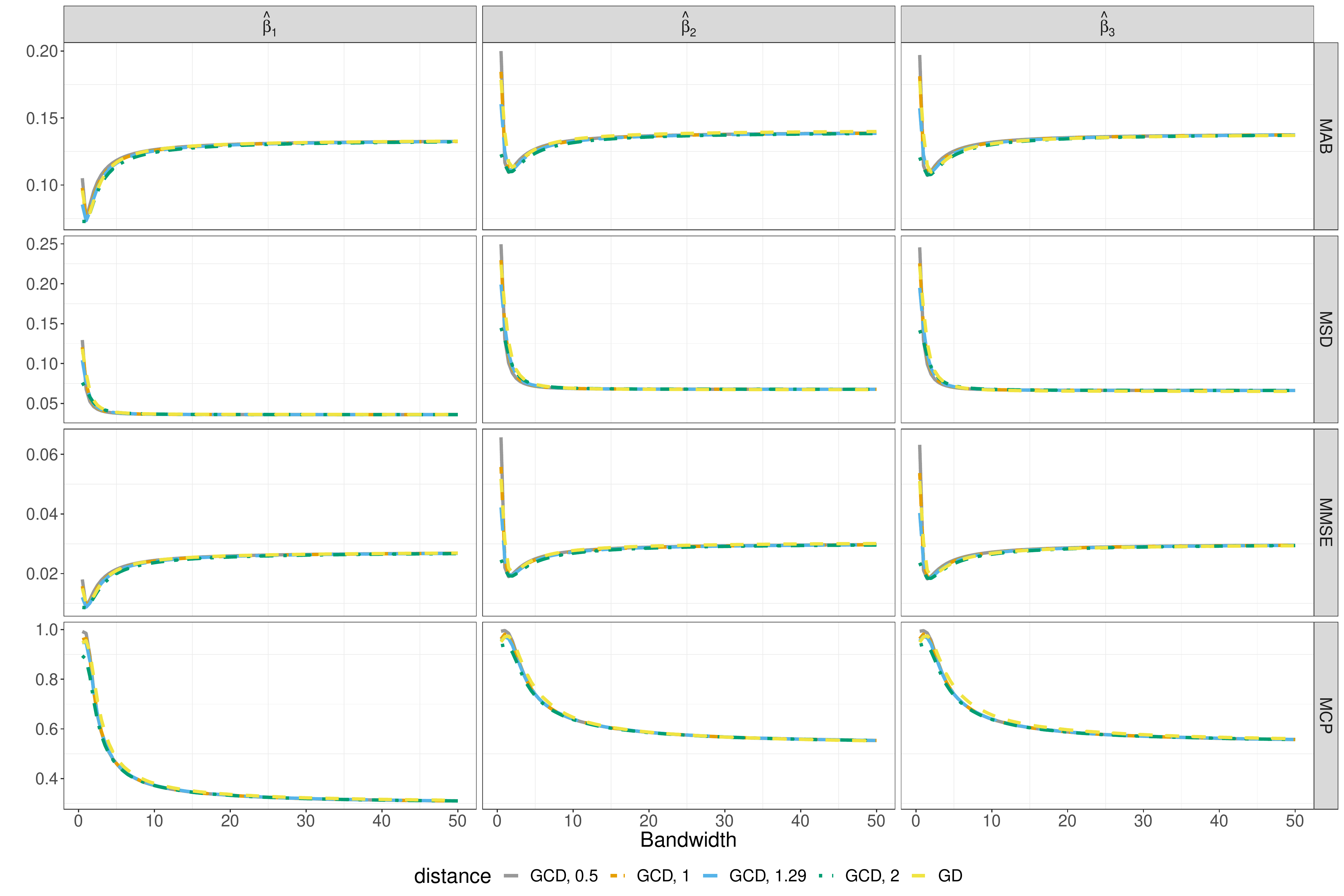}
	\caption{Performance measures for the geographically weighted Cox models
fitted using different distance/threshold combinations when the covariate
effects are spatially varying as in~\eqref{eq:alter1}.}
	\label{fig:alter1plot}
\end{figure}

We first consider a scenario where the vector of coefficients depend on the
geospatial location of the 64 county centroids (latitude, longitude). Same as
before, the same 7 models are fitted and 1000 replicates of simulation are
performed. For data generation, to simulate smooth variation over adjacent
locations, for county $\ell$, $\ell=1,\ldots, 64$, we let the $\beta$ vector for
county $\ell$ be the transpose of
\begin{equation}\label{eq:alter1}
	(0.7, 0.5, -0.8) + 0.15 \times (\text{latitude}_\ell-\bar{\text{latitude}}+
	\text{longitude}_\ell - \bar{\text{longitude}}).
\end{equation}
\begin{table}[tbp]
\centering 
\caption{Performance of local, global, and best selected geographically
weighted Cox regressions when covariate effects vary spatially according
to~\eqref{eq:alter1}.}
\label{tab:alter1}
\begin{tabular}{lccccc}
\toprule
	Model & Parameter & MAB & MSD & MMSE & MCP\\
	\midrule 
	Local & $\beta_1$ &  0.303 & 0.400 & 0.171 & 0.946    \\
	      & $\beta_2$ &  0.607 & 1.255 & 1.806 & 0.946      \\
          & $\beta_3$ &  0.571 & 1.007 & 1.085 & 0.945   \\ [.5ex]
	Global & $\beta_1$ & 0.134 & 0.036 & 0.027 & 0.298    \\
	      & $\beta_2$ &  0.141 & 0.067 & 0.031 & 0.528   \\
          & $\beta_3$ &   0.139 & 0.065 & 0.030 & 0.537  \\ [.5ex]
	GD Weighted, $h=1$ & $\beta_1$ & 0.079 & 0.088 & 0.010 & 0.956     \\
	      & $\beta_2$ &   0.139 & 0.170 & 0.031 & 0.971        \\
          & $\beta_3$ &    0.138 & 0.168 & 0.031 & 0.971     \\ [.5ex]
	GCD Weighted, $d_l=0.5, h=0.5$ & $\beta_1$ &  0.105 & 0.129 & 0.018 & 0.992  

\\
	      & $\beta_2$ &    0.200 & 0.250 & 0.066 & 0.993    \\
          & $\beta_3$ &   0.197 & 0.246 & 0.063 & 0.993 \\[.5ex]
	GCD Weighted, $d_l=1, h=1$ & $\beta_1$ & 0.074 & 0.076 & 0.009 & 0.964      
\\
	      & $\beta_2$ &  0.125 & 0.146 & 0.025 & 0.982     \\
          & $\beta_3$ &   0.123 & 0.144 & 0.024 & 0.982    \\ [.5ex]
	GCD Weighted, $d_l=1.29, h=1$ & $\beta_1$ & 0.073 & 0.076 & 0.009 & 0.948  \\
	      & $\beta_2$ &  0.125 & 0.146 & 0.025 & 0.970   \\
          & $\beta_3$ &   0.122 & 0.144 & 0.024 & 0.972 \\ [.5ex]
	GCD Weighted, $d_l=2, h=1$ & $\beta_1$ & 0.073 & 0.068 & 0.009 & 0.880 \\
	      & $\beta_2$ &   0.115 & 0.130 & 0.021 & 0.938        \\
          & $\beta_3$ &     0.113 & 0.128 & 0.021 & 0.941     \\
\bottomrule
\end{tabular}
\end{table}
The range for each true coefficient is around 0.72, and censoring rates in the
generated 64,000 blocks of data range from 16.7\% to 90.9\%.
Again, a visualization of the performance measures is given in
Figure~\ref{fig:alter1plot}. It can be seen that the performance of great
circle distance based models is influenced by the threshold $d_l$. At small
bandwidths, a small $d_l$ (0.5, for example) gives rise to a weighting scheme
such that the model produces unstable parameter estimates for each county,
while no weighting scheme based on a large $d_l$ (2, for example) can produce
parameter estimates that have reasonable coverage probability. As can be seen
from the graphs, there is a certain ``sweet zone'' of bandwidths, where
appropriate amount of information is taken from neighbors, such that the
geographically weighted parameter estimates have small MAB, MSD and MMSE. This
is where an appropriate balance between bias and variance is achieved. Similar
to Table~\ref{tab:null}, we recorded the most frequently selected bandwidth by
TIC, and report the corresponding performance measures, together with those for
the local and global Cox regressions, in Table~\ref{tab:alter1}. At $d_l=0.5$,
the optimal bandwidth selected by TIC for the great circle distance based
models is again 0.5, while at the other three $d_l$ values, 1 becomes the most
frequently chosen bandwidth. Given the performance of the graph distance based
model with $h=1$, we see it has better estimation than the great circle
distance based model with $d_l=0.5$, comparable performance to those with
$d_l=1$ or $d_l=1.29$,  and better coverage than that with $d_l=2$.

\begin{figure}[tbp]
	\centering
	\includegraphics[width = .8\textwidth]{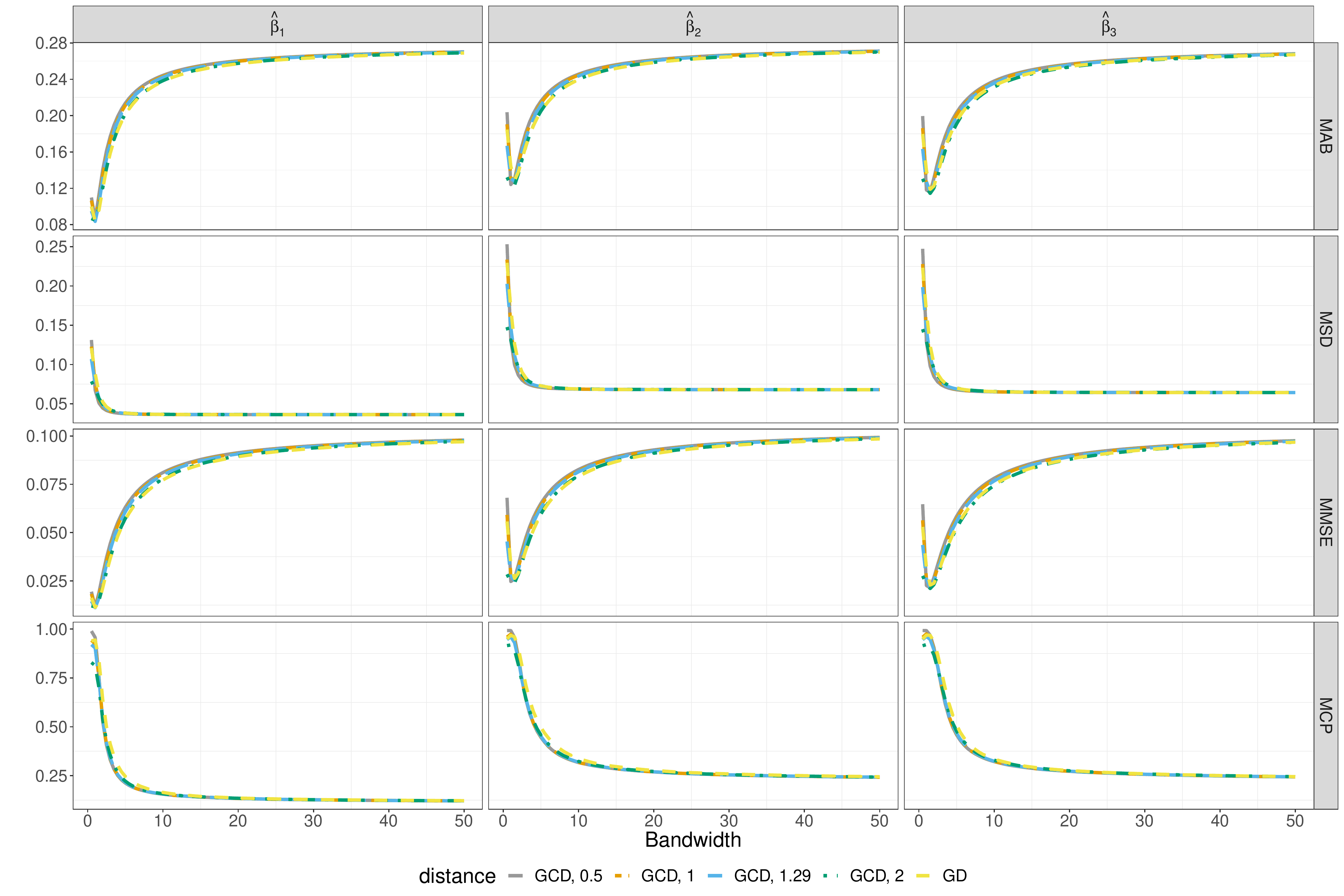}
	\caption{Performance measures for the geographically weighted Cox models
fitted using different distance/threshold combinations when the covariate
effects are spatially varying as in~\eqref{eq:alter2}.
}
\label{fig:alter2plot}
\end{figure}
\begin{table}[tbp]
\centering 
\caption{Performance of local, global, and best selected geographically
weighted Cox regressions when covariate effects vary spatially according
to~\eqref{eq:alter2}.}
\label{tab:alter2}
\begin{tabular}{lccccc}
\toprule
	Model & Parameter & MAB & MSD & MMSE & MCP\\
	\midrule 
	Local & $\beta_1$ & 0.308 & 0.411 & 0.181 & 0.945    \\
	      & $\beta_2$ &  0.692 & 1.562 & 3.336 & 0.947    \\
          & $\beta_3$ &  0.611 & 1.266 & 1.914 & 0.947  \\ [.5ex]
	Global & $\beta_1$ &  0.277 & 0.036 & 0.103 & 0.114 \\
	      & $\beta_2$ &  0.278 & 0.069 & 0.104 & 0.226  \\
          & $\beta_3$ &  0.276 & 0.066 & 0.103 & 0.230 \\ [.5ex]
	GD Weighted, $h=1$ & $\beta_1$ & 0.084 & 0.088 & 0.011 & 0.944 \\
	      & $\beta_2$ &  0.144 & 0.171 & 0.034 & 0.969  \\
          & $\beta_3$ &  0.140 & 0.168 & 0.032 & 0.969 \\ [.5ex]
	GCD Weighted, $d_l=0.5, h=0.5$ & $\beta_1$ & 0.110 & 0.131 & 0.019 & 0.990\\
	      & $\beta_2$ &  0.204 & 0.253 & 0.068 & 0.993  \\
          & $\beta_3$ &  0.200 & 0.247 & 0.065 & 0.993 \\[.5ex]
	GCD Weighted, $d_l=1, h=1$ & $\beta_1$ & 0.085 & 0.075 & 0.011 & 0.925 \\
	      & $\beta_2$ & 0.132 & 0.147 & 0.028 & 0.974 \\
          & $\beta_3$ & 0.127 & 0.144 & 0.026 & 0.975  \\ [.5ex]
	GCD Weighted, $d_l=1.29, h=1$ & $\beta_1$ & 0.083 & 0.076 & 0.011 & 0.903 \\
	      & $\beta_2$ & 0.131 & 0.147 & 0.028 & 0.961 \\
          & $\beta_3$ & 0.127 & 0.145 & 0.026 & 0.962 \\ [.5ex]
	GCD Weighted, $d_l=2, h=1$ & $\beta_1$ & 0.085 & 0.069 & 0.012 & 0.813 \\
	      & $\beta_2$ & 0.124 & 0.131 & 0.025 & 0.919   \\
          & $\beta_3$ & 0.121 & 0.130 & 0.024 & 0.921     \\
\bottomrule
\end{tabular}
\end{table}
In another scenario, instead of having the true covariate effects depend on
geographical locations, we generate the true parameter vectors based on graph
distance. The county St. Charles is selected as baseline with true parameter
vector $(0.7, 0.5, -0.8)^\top$. It is chosen as it is located among a cluster of
relatively small counties, so that geographically close counties can have a
large between county graph distance. For another county $\ell$, the parameter
vector is the transpose of
\begin{equation}\label{eq:alter2}
\begin{split}
		&(0.7, 0.5, -0.8) \\
		&+ 0.12\times (\text{the graph distance between county
$\ell$ and St. Charles } \\
&- \text{mean of graph distance between all other counties and St. Charles})
\end{split}.
\end{equation}
The produced coefficients for each covariate have a range of 1.08. Censoring
rates in generated data blocks range from 12.9\% to 94.3\%.
The corresponding results are presented in Figure~\ref{fig:alter2plot} and
Table~\ref{tab:alter2}. Again, similar to in Figure~\ref{fig:alter1plot}, the
tradeoff between bias and variance is clear. The performance of the graph
distance based models is fairly robust - MAB, MSD and MMSE of parameter
estimates are greatly reduced, while the MCP is maintained at around 0.95.
The great circle distance based models, however, perform rather
differently when different $d_l$ values are used. At $d_l=1.29$ and $d_l=2$,
their performance is fairly similar to the graph distance based model.
Indeed, while a good balance between bias and variance can also be 
found with great circle distance based models with an appropriate threshold 
$d_l$ and bandwidth $h$, the graph distance based models bypass the need to 
select a threshold and only require selection of $h$.

\section{Real Data Analysis}\label{sec:realdata}

We consider the prostate cancer data for Louisiana from the SEER Program. As in
\cite{onicescu2018bayesian}, we exclude observations that have unknown ending
statuses or unknown survival times, resulting in 1,277 complete observations.
Descriptive statistics for the dataset are presented in
Table~\ref{tab:demo}. We consider the same three covariates as in
data simulation, i.e., \textsf{Age}, \textsf{Black}, and \textsf{Married}.

\begin{figure}[tbp]
\centering
\includegraphics[width = .8\linewidth]{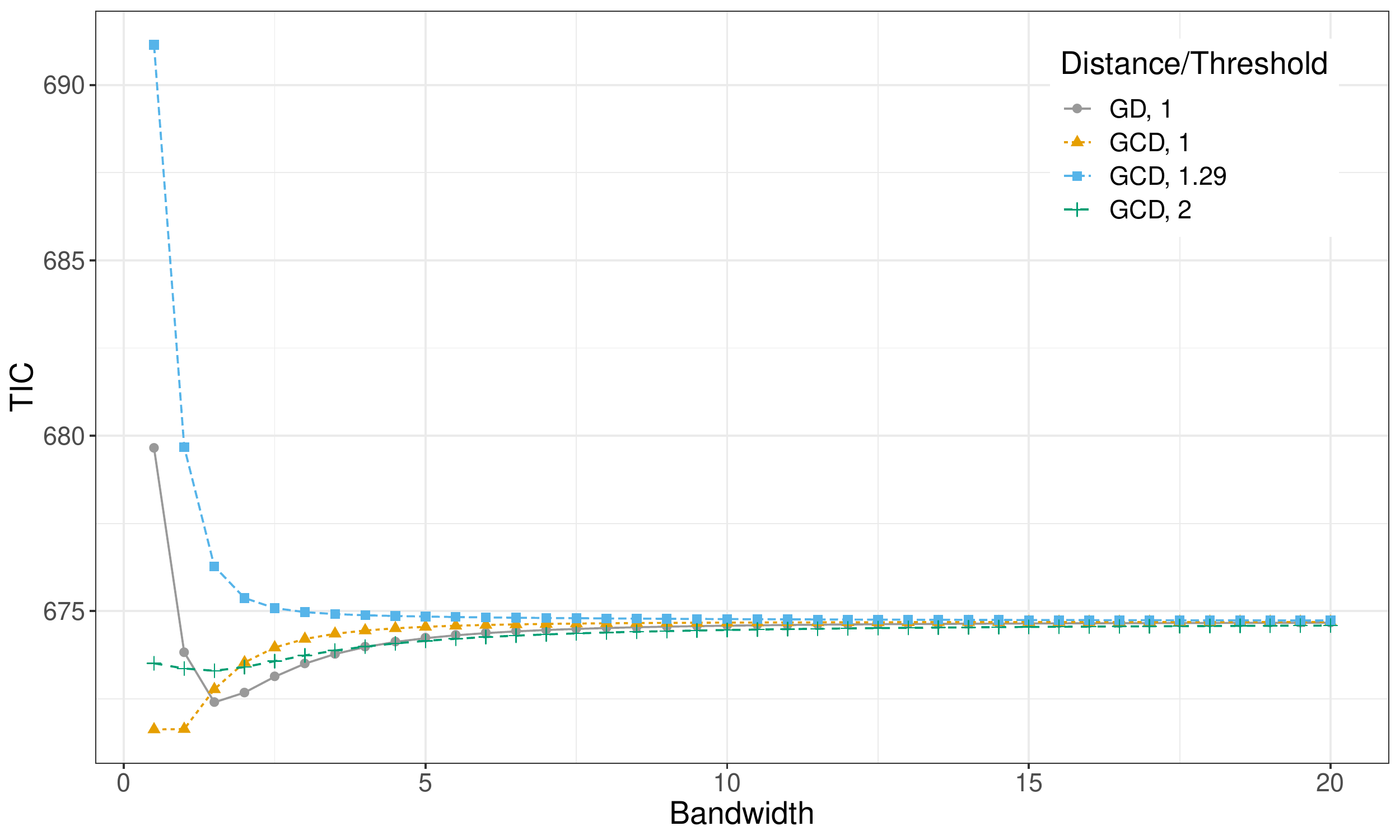}
\caption{TICs calculated for geographically weighted Cox models plotted against
their respective bandwidths.}\label{fig:tic}
\end{figure}

We use the proposed weighted technique to obtain parameter estimates for each of
the 64 counties. In addition to utilizing the graph distance, we also use the
matrix of great circle distance between the county centroids for comparison. The
same normalization is done to make the two distances comparable.  Based on the
simulation results, setting $d_l=0.5$ for great circle distance tends to produce
unstable parameter estimates for each county. Therefore, only 1, 1.29 and 2 are
considered for candidate $d_l$ values. Bandwidths $h \in \{0.5, \ldots, 20\}$
are considered, as we have seen in simulation studies that model performances
are rather stable beyond $h=20$. For each $h$, we calculated the TIC as in
\eqref{eq:TIC_b}. As much as we want to make performance comparisons of the
obtained parameter estimates and local estimates, such comparison is impossible,
as four of the 64 counties had less than three observations, and local estimates
for these counties cannot be obtained. Nevertheless, we are able to calculate
the score vectors using per-county data, and calculate the TIC for bandwidth
selection. The TICs are plotted against their corresponding bandwidths in
Figure~\ref{fig:tic}. The TIC for the graph distance-based model is minimized at
around $h=1.5$ with a value of 672.4, where we have a good balance of bias and
variance. When $h$ further increases, the TIC also increases, as larger
bandwidths incur larger within-county biases.

For great circle distance based model at $d_l=1$, $h=0.5$ is selected, with a
minimized TIC value of 671.6. When the threshold $d_l$ is further increased to
1.29, the TIC values show a monotonically decreasing trend, and favors the
largest bandwidth, 20, with corresponding TIC value 674.7. Finally, at $d_l=2$,
the minimum TIC is attained at $h=1.5$, with value 673.3. In addition, as there
are counties with no events or only few observations, the local regressions
could not be performed, but the global Cox regression can be fitted. The
resulting TIC has a value of 674.7. Again, the performance of the great circle
distance based models depends on the choice of threshold, which further
influences the result of bandwidth selection. The graph distance, however,
induces a natural weighting scheme that remains robust and provides credible
estimation results.

\begin{figure}[tbp]
\centering
    \includegraphics[width = .8\textwidth]{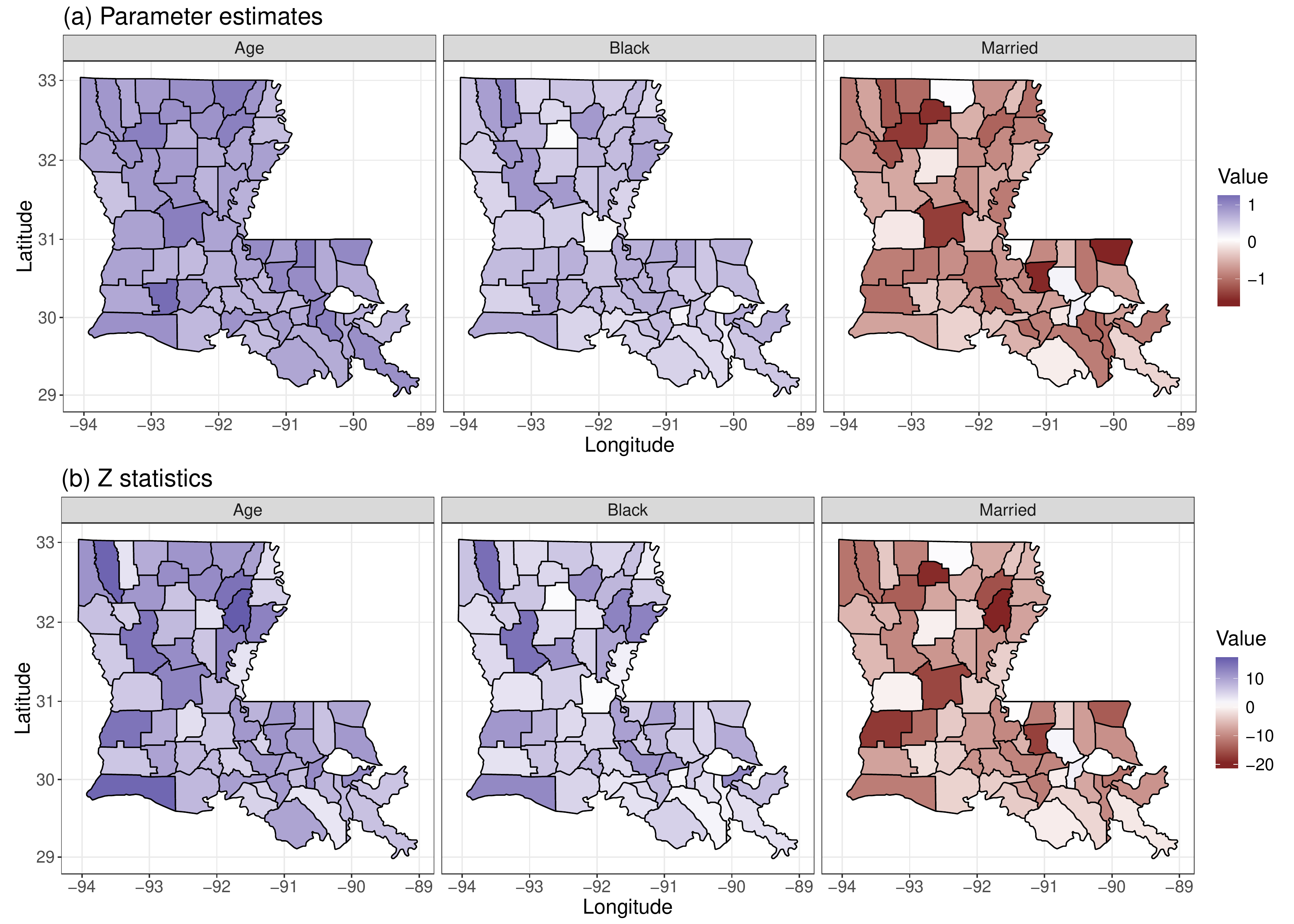}
    \caption{Parameter estimates (panel (a)), and their corresponding $Z$
    statistics (panel (b)) obtained from the geographically weighted Cox model
    for Louisiana counties using $h = 1.5$.}\label{fig:betaZ}
\end{figure}

The parameter estimates of the final model based on graph distance and with
$h=1.5$ corresponding to each county are plotted in panel~{(a)} of
Figure~\ref{fig:betaZ}, together with their corresponding $Z$ statistics
($\hat{\beta}_{\ell, i} / \text{SE}(\hat{\beta}_{\ell, i})$ for $\ell=1,\ldots,
64$ and $i=1,2, 3$) plotted in panel~{(b)}. It can be easily observed that the
effects of all three covariates are geographically varying. All parameters for
\textsf{Age} are positive, indicating that increase in age brings higher hazard
in all counties, which is in accordance with our intuition. Compared with other
races, black males have been observed to have higher hazard, which could be of
interest for studies on racial disparities in healthcare outcomes. In 61 of all
64 counties, married males have lower hazard for dying of prostate cancer. In
the other three counties, the increasing effect is minor
($\hat{\beta}_{\textsf{Marriage}} = 0.09, 0.12$ and $ 0.03$, and neither
significantly different from 0).

\section{Discussion}\label{sec:disc}

In this paper, we proposed a geographically weighted Cox regression model to
allow for varying coefficients at the local or subregional level, and
corresponding partial likelihood based selection criterion for choosing the
required bandwidth. In the simulation study, under the null scenario where there
is no geographical variation in the covariate effects, we find that our method
performed better than the local stratification estimation, which only used the
observations within each county, and has similar performance with models that
use the great circle distance weighted scheme and global estimation scheme. When
simulating under scenarios in which there was spatial variation in covariate
effects, the proposed selection criterion effectively selected the model with
parameter estimates that have small within county variance yet still maintain
high average coverage probability. In comparison, great circle distance based
models perform differently when the threshold is set different values. The graph
distance based models, however, do not suffer from this issue. The findings from
the survival analysis of SEER prostate cancer patients are clearly appealing.
Despite the sparsity of data, we are able to obtain parameter estimates for all
64 counties. Compared to the great circle distance, the graph distance, together
with its natural threshold of 1 in defining being ``close enough'', provides a
robust weighting scheme that produces models which achieve an appropriate
balance between the bias and variance of parameter estimates for each location.

For future work, we will concentrate more on the following aspects. First, our
bandwidth selection method is based on the TIC using the partial likelihood.
Other criteria that focus on prediction, such as cross-validation methods, are
worth investigating. Second, we assume all the regression coefficients are
spatially varying. In practice, some regression coefficients may not be
spatially varying. Identifying which coefficients are spatially varying, and
estimating the Cox model with both spatially varying coefficients and spatially
constant coefficients, such as in \cite{mei2004note} for linear regression, are
another two important areas for future exploration. The implementation of
standard tests for the proportional hazards assumption, such as that in
\cite{grambsch1994}, in the context of spatially varying coefficients, is also
worth investigating. \blue{In addition,} when there is a large number of
covariates, it
may be of interest to develop a geographically penalized Cox regression model
for variable selection. \blue{Finally, in this work, each covariate is weighed
using
the same value of bandwidth. Multiscale GWR, which allows each covariate to be
weighed differently and allows more model flexibility, has been proposed
recently by \cite{fotheringham2017multiscale}, and implemented in
\textsf{Python} by
\cite{oshan2019mgwr}. \cite{yu2019} re-framed the multiscale GWR as an additive
model and enabled inference for parameters estimates. 
Multiscale geographically weighted survival models, such as Cox model and the
accelerated failure time model, are also devoted to future research.}



\begin{thebibliography}{}

\bibitem[\protect\citeauthoryear{Akaike}{Akaike}{1973}]{akaike1973information}
Akaike, H. (1973).
\newblock Information theory and an extension of the maximum likelihood
  principle.
\newblock In B.~N. Petrov and F.~Csaki (Eds.), {\em Second International
  Symposium on Information Theory}, pp.\  267--281. Akad\'{e}miai Kiado.

\bibitem[\protect\citeauthoryear{Banerjee, Carlin, and Gelfand}{Banerjee
  et~al.}{2014}]{banerjee2014hierarchical}
Banerjee, S., B.~P. Carlin, and A.~E. Gelfand (2014).
\newblock {\em Hierarchical Modeling and Analysis for Spatial Data}.
\newblock CRC press.

\bibitem[\protect\citeauthoryear{Banerjee and Dey}{Banerjee and
  Dey}{2005}]{banerjee2005semiparametric}
Banerjee, S. and D.~K. Dey (2005).
\newblock Semiparametric proportional odds models for spatially correlated
  survival data.
\newblock {\em Lifetime Data Analysis\/}~{\em 11\/}(2), 175--191.

\bibitem[\protect\citeauthoryear{Banerjee, Wall, and Carlin}{Banerjee
  et~al.}{2003}]{banerjee2003frailty}
Banerjee, S., M.~M. Wall, and B.~P. Carlin (2003).
\newblock Frailty modeling for spatially correlated survival data, with
  application to infant mortality in {M}innesota.
\newblock {\em Biostatistics\/}~{\em 4\/}(1), 123--142.

\bibitem[\protect\citeauthoryear{Bhattacharyya and Bickel}{Bhattacharyya and
  Bickel}{2014}]{bhattacharyya2014community}
Bhattacharyya, S. and P.~J. Bickel (2014).
\newblock Community detection in networks using graph distance.
\newblock {\em arXiv preprint arXiv:1401.3915\/}.

\bibitem[\protect\citeauthoryear{Brunsdon, Fotheringham, and Charlton}{Brunsdon
  et~al.}{1996}]{brunsdon1996geographically}
Brunsdon, C., A.~S. Fotheringham, and M.~E. Charlton (1996).
\newblock Geographically weighted regression: a method for exploring spatial
  nonstationarity.
\newblock {\em Geographical Analysis\/}~{\em 28\/}(4), 281--298.

\bibitem[\protect\citeauthoryear{Brunsdon, Fotheringham, and Charlton}{Brunsdon
  et~al.}{1998}]{brunsdon1998geographically}
Brunsdon, C., S.~Fotheringham, and M.~Charlton (1998).
\newblock Geographically weighted regression-modelling spatial
  non-stationarity.
\newblock {\em Journal of the Royal Statistical Society: Series D (The
  Statistician)\/}~{\em 47\/}(3), 431--443.

\bibitem[\protect\citeauthoryear{Cox}{Cox}{1972}]{cox1972}
Cox, D.~R. (1972).
\newblock Regression models and life-tables.
\newblock {\em Journal of the Royal Statistical Society. Series B
  (Methodological)\/}~{\em 34\/}(2), 187--220.

\bibitem[\protect\citeauthoryear{Cox}{Cox}{1975}]{cox1975partial}
Cox, D.~R. (1975).
\newblock Partial likelihood.
\newblock {\em Biometrika\/}~{\em 62\/}(2), 269--276.

\bibitem[\protect\citeauthoryear{Fan, Lin, Zhou, et~al.}{Fan
  et~al.}{2006}]{fan2006local}
Fan, J., H.~Lin, Y.~Zhou, et~al. (2006).
\newblock Local partial-likelihood estimation for lifetime data.
\newblock {\em The Annals of Statistics\/}~{\em 34\/}(1), 290--325.

\bibitem[\protect\citeauthoryear{Fleming and Harrington}{Fleming and
  Harrington}{1991}]{fleming1991}
Fleming, T.~R. and D.~P. Harrington (1991).
\newblock {\em Counting Processes and Survival Analysis}.
\newblock New York: Wiley.

\bibitem[\protect\citeauthoryear{Fotheringham, Yang, and Kang}{Fotheringham
  et~al.}{2017}]{fotheringham2017multiscale}
Fotheringham, A.~S., W.~Yang, and W.~Kang (2017).
\newblock Multiscale geographically weighted regression {(mgwr)}.
\newblock {\em Annals of the American Association of Geographers\/}~{\em
  107\/}(6), 1247--1265.

\bibitem[\protect\citeauthoryear{Gelfand, Kim, Sirmans, and Banerjee}{Gelfand
  et~al.}{2003}]{gelfand2003spatial}
Gelfand, A.~E., H.-J. Kim, C.~Sirmans, and S.~Banerjee (2003).
\newblock Spatial modeling with spatially varying coefficient processes.
\newblock {\em Journal of the American Statistical Association\/}~{\em
  98\/}(462), 387--396.

\bibitem[\protect\citeauthoryear{Grambsch and Therneau}{Grambsch and
  Therneau}{1994}]{grambsch1994}
Grambsch, P.~M. and T.~M. Therneau (1994).
\newblock Proportional hazards tests and diagnostics based on weighted
  residuals.
\newblock {\em Biometrika\/}~{\em 81\/}(3), 515--526.

\bibitem[\protect\citeauthoryear{Hijmans}{Hijmans}{2017}]{Rpkg:geosphere}
Hijmans, R.~J. (2017).
\newblock {\em {geosphere}: Spherical Trigonometry}.
\newblock {R} package version 1.5-7.

\bibitem[\protect\citeauthoryear{Hu}{Hu}{2017}]{hu2017spatial}
Hu, G. (2017).
\newblock {\em Spatial Statistics and Its Applications in Biostatistics and
  Environmental Statistics}.
\newblock Ph.\ D. thesis, The Florida State University.

\bibitem[\protect\citeauthoryear{Hu and Huffer}{Hu and
  Huffer}{2019}]{hu2018modified}
Hu, G. and F.~Huffer (2019).
\newblock Modified {K}aplan--{M}eier estimator and {N}elson--{A}alen estimator
  with geographical weighting for survival data.
\newblock {\em Geographical Analysis\/}.
\newblock Forthcoming.

\bibitem[\protect\citeauthoryear{Li, Hanson, and Zhang}{Li
  et~al.}{2015}]{li2015spatial}
Li, L., T.~Hanson, and J.~Zhang (2015).
\newblock Spatial extended hazard model with application to prostate cancer
  survival.
\newblock {\em Biometrics\/}~{\em 71\/}(2), 313--322.

\bibitem[\protect\citeauthoryear{Lu, Brunsdon, Charlton, and Harris}{Lu
  et~al.}{2019}]{lu2019response}
Lu, B., C.~Brunsdon, M.~Charlton, and P.~Harris (2019).
\newblock A response to 'a comment on geographically weighted regression with
  parameter-specific distance metrics'.
\newblock {\em International Journal of Geographical Information
  Science\/}~{\em 33\/}(7), 1300--1312.

\bibitem[\protect\citeauthoryear{Mei, He, and Fang}{Mei
  et~al.}{2004}]{mei2004note}
Mei, C.-L., S.-Y. He, and K.-T. Fang (2004).
\newblock A note on the mixed geographically weighted regression model.
\newblock {\em Journal of Regional Science\/}~{\em 44\/}(1), 143--157.

\bibitem[\protect\citeauthoryear{M{\"u}ller, Szymanski, Knop, and
  Trinajsti{\'c}}{M{\"u}ller et~al.}{1987}]{muller1987algorithm}
M{\"u}ller, W., K.~Szymanski, J.~Knop, and N.~Trinajsti{\'c} (1987).
\newblock An algorithm for construction of the molecular distance matrix.
\newblock {\em Journal of Computational Chemistry\/}~{\em 8\/}(2), 170--173.

\bibitem[\protect\citeauthoryear{Murakami, Lu, Harris, Brunsdon, Charlton,
  Nakaya, and Griffith}{Murakami et~al.}{2019}]{murakami2019importance}
Murakami, D., B.~Lu, P.~Harris, C.~Brunsdon, M.~Charlton, T.~Nakaya, and D.~A.
  Griffith (2019).
\newblock The importance of scale in spatially varying coefficient modeling.
\newblock {\em Annals of the American Association of Geographers\/}~{\em
  109\/}(1), 50--70.

\bibitem[\protect\citeauthoryear{Nakaya, Fotheringham, Brunsdon, and
  Charlton}{Nakaya et~al.}{2005}]{nakaya2005geographically}
Nakaya, T., A.~S. Fotheringham, C.~Brunsdon, and M.~Charlton (2005).
\newblock Geographically weighted {P}oisson regression for disease association
  mapping.
\newblock {\em Statistics in Medicine\/}~{\em 24\/}(17), 2695--2717.

\bibitem[\protect\citeauthoryear{Onicescu and Lawson}{Onicescu and
  Lawson}{2018}]{onicescu2018bayesian}
Onicescu, G. and A.~B. Lawson (2018).
\newblock Bayesian cure-rate survival model with spatially structured
  censoring.
\newblock {\em Spatial Statistics\/}~{\em 28}, 352--364.

\bibitem[\protect\citeauthoryear{Oshan, Wolf, Fotheringham, Kang, Li, and
  Yu}{Oshan et~al.}{2019}]{oshan2019comment}
Oshan, T., L.~J. Wolf, A.~S. Fotheringham, W.~Kang, Z.~Li, and H.~Yu (2019).
\newblock A comment on geographically weighted regression with
  parameter-specific distance metrics.
\newblock {\em International Journal of Geographical Information
  Science\/}~{\em 33\/}(7), 1289--1299.

\bibitem[\protect\citeauthoryear{Oshan, Li, Kang, Wolf, and Fotheringham}{Oshan
  et~al.}{2019}]{oshan2019mgwr}
Oshan, T.~M., Z.~Li, W.~Kang, L.~J. Wolf, and A.~S. Fotheringham (2019).
\newblock {mgwr}: A {P}ython implementation of multiscale geographically
  weighted regression for investigating process spatial heterogeneity and
  scale.
\newblock {\em ISPRS International Journal of Geo-Information\/}~{\em 8\/}(6),
  269.

\bibitem[\protect\citeauthoryear{Takeuchi}{Takeuchi}{1976}]{Takeuchi:1976fb}
Takeuchi, K. (1976).
\newblock The distribution of information statistic and the criterion of the
  adequacy of a model.
\newblock {\em Suri-Kagaku (Mathematical Sciences)\/}~{\em 3}, 12--18.
\newblock (in Japanese).

\bibitem[\protect\citeauthoryear{Therneau}{Therneau}{2017}]{therneau2017}
Therneau, T.~M. (2017).
\newblock {\em A Package for Survival Analysis in \textsf{S}}.
\newblock version 2.41-3.

\bibitem[\protect\citeauthoryear{Tobler}{Tobler}{1970}]{tobler1970computer}
Tobler, W.~R. (1970).
\newblock A computer movie simulating urban growth in the {D}etroit region.
\newblock {\em Economic Geography\/}~{\em 46}, 234--240.

\bibitem[\protect\citeauthoryear{Verweij and van Houwelingen}{Verweij and van
  Houwelingen}{1995}]{verweij1995time}
Verweij, P.~J. and H.~C. van Houwelingen (1995).
\newblock Time-dependent effects of fixed covariates in {C}ox regression.
\newblock {\em Biometrics\/}~{\em 51\/}(4), 1550--1556.

\bibitem[\protect\citeauthoryear{Wang and Yao}{Wang and
  Yao}{2006}]{wang2006estimation}
Wang, Q. and L.~Yao (2006).
\newblock Estimation in varying-coefficient proportional hazard regression
  model.
\newblock {\em Metrika\/}~{\em 64\/}(3), 271--288.

\bibitem[\protect\citeauthoryear{White and Ghosh}{White and
  Ghosh}{2009}]{white2009stochastic}
White, G. and S.~K. Ghosh (2009).
\newblock A stochastic neighborhood conditional autoregressive model for
  spatial data.
\newblock {\em Computational Statistics \& Data Analysis\/}~{\em 53\/}(8),
  3033--3046.

\bibitem[\protect\citeauthoryear{Yu, Fotheringham, Li, Oshan, Kang, and
  Wolf}{Yu et~al.}{2019}]{yu2019}
Yu, H., A.~S. Fotheringham, Z.~Li, T.~Oshan, W.~Kang, and L.~J. Wolf (2019).
\newblock Inference in multiscale geographically weighted regression.
\newblock {\em Geographical Analysis\/}.
\newblock Forthcoming.

\bibitem[\protect\citeauthoryear{Zhang and Lawson}{Zhang and
  Lawson}{2011}]{zhang2011bayesian}
Zhang, J. and A.~B. Lawson (2011).
\newblock Bayesian parametric accelerated failure time spatial model and its
  application to prostate cancer.
\newblock {\em Journal of Applied Statistics\/}~{\em 38\/}(3), 591--603.

\end{thebibliography}

\end{document}